\documentclass[AMA,STIX1COL]{WileyNJD-v2}
\usepackage{moreverb}
\usepackage{subfigure} 
\usepackage{subcaption}
\usepackage{graphicx}

\newcommand\BibTeX{{\rmfamily B\kern-.05em \textsc{i\kern-.025em b}\kern-.08em
T\kern-.1667em\lower.7ex\hbox{E}\kern-.125emX}}

\articletype{Article Type}%

\received{<day> <Month>, <year>}
\revised{<day> <Month>, <year>}
\accepted{<day> <Month>, <year>}

\begin{document}

\title{Two-Dimensional Locally Adaptive Non-Hydrostatic Extension of Shallow Water Equations}

\author[1,2]{Kemal Firdaus*}

\author[1,2]{J\"orn Behrens}

\authormark{FIRDAUS and BEHRENS}

\address[1]{Department of Mathematics, Universit\"at Hamburg, Bundesstrasse 53-55, 20146 Hamburg, Germany}

\address[2]{Center for Earth System Research and Sustainability (CEN), Universit\"at Hamburg, Bundesstrasse 53-55, 20146 Hamburg, Germany}

\corres{*Kemal Firdaus, Department of Mathematics, Universit\"at Hamburg, Bundesstrasse 53-55, 20146 Hamburg, Germany.\\ \email{kemal.firdaus@uni-hamburg.de}}


\abstract[Abstract]{We introduce a two-dimensional non-hydrostatic model for shallow water wave dispersion. The model is based on a locally adapted application of a non-hydrostatic correction to the hydrostatic shallow water equations (SWE) in a predictor-corrector scheme. Applying the non-hydrostatic correction uniformly to the entire domain demands a high computational cost, since an elliptic system of equations needs to be solved for the correction terms. We demonstrate that by determining the area where the non-hydrostatic effects are significant, and applying the correction only locally, the computational effort can be reduced by approximately 40\% without sacrificing accuracy in tsunami-like scenarios. As indicators for the non-hydrostatic effect, we use the ratio between total water depth and surface elevation, as well as horizontal velocity norms. Results are shown for several well-known test cases, including wave trains over a semi-circular shoal, static, and moving bottom tsunami-like wave propagation.
}


\maketitle

\section{Introduction}\label{sec1}
 In some geophysical flow phenomena, such as landslides and slow earthquake-generated tsunamis, non-hydrostatic pressure is necessary to represent the physics of the phenomenon \cite{KIRBY2022101943, glimsdal2013}. This is due to a so-called dispersive effect, which is driven by the horizontal gradient of the non-hydrostatic pressure. To cover such an effect in moving bottom-generated waves, three-dimensional models, such as Navier-Stokes equations, have been successfully employed \cite{ABADIE2010779, AI20211, YUK2006927}. However, due to their three-dimensional nature, it is impractical to adopt them for large-scale ocean modelling. Preferably, depth-averaged models have been proposed, which involve higher derivatives or solving elliptic problems to represent the non-hydrostatic effects. Solving for the dispersive non-hydrostatic terms over the whole domain at each time step is still computationally intensive. In this study we develop a model that allows to solve the elliptic problems only locally, adapted to areas where non-hydrostatic effects might be crucial, thereby reducing the computational cost while preserving the accuracy.
 
 Among the depth-integrated non-hydrostatic models are the Boussinesq-type equations, which have been widely used to simulate dispersive waves. Various Boussinesq-type equations are derived based on the asymptotic expansion of the velocity potential with different expansion orders \cite{madsen1998, madsen2003, MADSEN2006487}. Such a model has been widely used in simulating geophysical flow, especially tsunami wave propagation \cite{FUHRMAN2009747, FANG2020101977, dutykh2013}. However, this approach involves higher-order mixed time-space derivatives, making it numerically and computationally challenging. 

Alternatively, the shallow water equations (SWE) have been widely used in the tsunami modelling community, and they form a purely hyperbolic, robust, and widely applicable model. However, it is limited by the hydrostatic assumption, rendering it unsuitable for dispersive wave propagation. Some studies have extended SWE by including the non-hydrostatic pressure terms. To achieve a purely depth-averaged model, a vertical relation of the non-hydrostatic pressure and vertical velocity needs to be defined. Some studies assume a linear profile \cite{stelling2003, walters2005, yamazaki2009}, while others employ a quadratic relation \cite{Jeschke2017,wang2020, DEMPWOLFF2024}, where the vertical velocity is assumed to be linear. The latter pressure relation is proven to be equivalent to the Green-Naghdi equations \cite{Green_Naghdi_1976}, which are suitable for weakly dispersive waves. Such an extension can be solved with a projection method, frequently implemented as a predictor-corrector method, where the non-hydrostatic correction is applied to the hydrostatic SWE being the predictor. 

The projection method is one of the key ingredients of the locally adaptive model, as it allows for computing the correction terms locally. Another ingredient is a proper criterion to define the corrected region. One possible way is to define the criterion based on the hydrostatic solution. In one-dimensional settings, a locally adaptive model was proposed by simply taking the norm of the hydrostatic solution as the criterion \cite{firdaus2026}. This approach has shown comparable results with the global model, while saving more than half the computational time. 

This work extends the applicability of such a locally adaptive model to two-dimensional settings. We introduce the two-dimensional modified quadratic pressure relation that is solvable with a projection method without requiring to neglect terms. This is an extension of the one-dimensional form \cite{firdaus2025}. The adaptivity criterion is defined based on the magnitude of the surface elevation-fluid depth ratio, along with the horizontal velocity. Both quantities can be extracted from the purely hydrostatic predictor step. We apply our model to various test cases based on  experimental measurements. First, we simulate periodic waves over a semi-circular shoal. Then, we apply it to two tsunami-based experiments, involving a static and a moving bottom.

\section{Mathematical Model}\label{sec2}
This study employs a two-dimensional depth-averaged non-hydrostatic extension of the SWE. To derive this model, we begin with the three-dimensional Euler equations of motion
\begin{equation}\label{eq:euler_1}
    \boldsymbol{\nabla}_3\cdot \boldsymbol{V}= 0,
\end{equation}
\begin{equation}\label{eq:euler_2}
    \partial_t\boldsymbol{V}+\boldsymbol{\nabla}_3\cdot(\boldsymbol{V}\boldsymbol{V}^T)=-\frac{1}{\rho}\boldsymbol{\nabla}_3P-g\boldsymbol{E}_z,
\end{equation}
where $\boldsymbol{\nabla}_3 = (\partial_x,\partial_y,\partial_z)^T$ is the three-dimensional spatial gradient operator, $\boldsymbol{V}=(U,V,W)^T=(\boldsymbol{U},W)$ is the velocity in $x,y,z$-direction respectively, and $\boldsymbol{E}_z$ denotes the unit vector in $z$-direction. Moreover, $\rho$ denotes the fluid density, $P$ is the pressure, and $g$ is the gravitational acceleration. Kinematic boundary conditions at the fluid surface ($z=\eta$) and bottom ($z=-d$), along with the pressure value at the surface,
\begin{equation}\label{eq:kinematic_surface}
    W|_{z=\eta}:=W(x,y,\eta,t) = \partial_t\eta+\begin{pmatrix}U|_{z=\eta}\\V|_{z=\eta}\end{pmatrix}\cdot\boldsymbol{\nabla}\eta,
\end{equation}
\begin{equation}\label{eq:kinematic_bottom}
    W|_{z=-d}:=W(x,y,-d,t) = -\partial_td-\begin{pmatrix}U|_{z=-d}\\V|_{z=-d}\end{pmatrix}\cdot\boldsymbol{\nabla}d,
\end{equation}
\begin{equation}\label{eq:pressure_surface}
    P|_{z=\eta}=0,
\end{equation}
complete this set of equations. The key to obtaining a non-hydrostatic extension of the SWE is to consider the non-hydrostatic pressure, which can be achieved by splitting the pressure terms into hydrostatic and non-hydrostatic parts, such that $P = P^{hy}+P^{nh}$. 

Integrating Equations \eqref{eq:euler_1} and \eqref{eq:euler_2} over the fluid depth $h=\eta+d$ with the boundary conditions (Equations \eqref{eq:kinematic_surface}-\eqref{eq:pressure_surface}) and applying Leibniz integration rule, we obtain
\begin{equation}\label{eq:swe_mass}
    h_t+\nabla\cdot(h\boldsymbol{u})=0,
\end{equation}
\begin{equation}\label{eq:swe_momentum_horizontal}
    (h\boldsymbol{u})_t+\nabla\cdot(h\boldsymbol{u}\otimes\boldsymbol{u}+\frac{g}{2}h^2\boldsymbol{I}_2) = gh\nabla d+\frac{1}{\rho}(P^{nh}|_{z=-d}\nabla d-\nabla(hp^{nh})),
\end{equation}
\begin{equation}\label{eq:swe_momentum_vertical}
    (hw)_t+\nabla\cdot(h\boldsymbol{u}w) = \frac{1}{\rho}P^{nh}|_{z=-d},
\end{equation}
with the lower-case letters denote the depth-averaged values. We call Equation \eqref{eq:swe_mass} the mass conservation equation and Equations \eqref{eq:swe_momentum_horizontal} and \eqref{eq:swe_momentum_vertical} the horizontal and vertical momentum balance equations, respectively. To achieve a purely averaged system, a divergent constraint is derived by assuming a linear vertical velocity relation and approximating the horizontal velocities $\boldsymbol{U}$ with their depth-averaged value $\boldsymbol{u}$, which leads us to 
\begin{equation}\label{eq:swe_constraint}
    2hw +h\boldsymbol{u}\cdot\nabla(2d-h)+2h\partial_td = -h\nabla\cdot(h\boldsymbol{u}).
\end{equation}

Note that we still need to define the relation between the non-hydrostatic pressure at the fluid bottom $P^{nh}|_{z=-d}$ with the averaged value $p^{nh}$. We consider two types of relations: linear and quadratic. The former relation is popular among studies on SWE non-hydrostatic extension, including those with a multilayer approach \cite{CUI20141, POPINET2020109609, magdalena2020}. This relation is based on a linear pressure profile along the vertical axis direction, yielding 
\begin{equation}\label{eq:pressure_linear}
    P^{nh}|_{z=-d}=2p^{nh}.
\end{equation}
The quadratic relation, on the other hand, was initially introduced by \cite{Jeschke2017}, where it was also proven to be equivalent to Boussinesq-type equations, specifically the Green-Naghdi equations. This model has also been developed to model bottom-generated waves while avoiding the previously required simplification in one-dimensional form \cite{firdaus2025}. With a similar approach, an alternative form of the two-dimensional pressure relation can be achieved by making use of the non-conservative form of the horizontal momentum 
\begin{equation}\label{eq:swe_momentum_nonconservative}
    \boldsymbol{u}_t+\nabla\cdot(\boldsymbol{u}\otimes\boldsymbol{u}+\frac{g}{2}h\boldsymbol{I}_2) = g\nabla d+\frac{1}{\rho h}(P^{nh}|_{z=-d}\nabla d-\nabla(hp^{nh})),
\end{equation}
yielding 
\begin{equation}\label{eq:pressure_quadratic}
    P^{nh}|_{z=-d} = \frac{6}{4+\nabla d\cdot\nabla d}p^{nh}+\frac{\nabla d}{4+\nabla d\cdot\nabla d}\cdot\nabla(\tilde{h}p^{nh})+\phi,
\end{equation}
where $\phi = \frac{\rho h}{4+\nabla d\cdot\nabla d}(g\nabla d\cdot\nabla \eta-\boldsymbol{u}\cdot\nabla(\nabla d)\cdot\boldsymbol{u}-d_{tt}-2\boldsymbol{u}\cdot\nabla d_t)$. This proposed relation can be directly solved with the projection method without any simplification, conserving its equivalence with the Green-Naghdi equations. Both the linear \eqref{eq:pressure_linear} and quadratic \eqref{eq:pressure_quadratic} relations complete the set of equations \eqref{eq:swe_mass}-\eqref{eq:swe_constraint}. 

\section{Numerical Method}\label{sec3}
We solve the previously described model with a projection method, which exploits the solution of the hydrostatic SWE as a predictor, followed by a correction by solving the remaining terms implicitly. We apply the correction for each Runge-Kutta step, instead of just for each timestep integration, to preserve the second-order in time accuracy. To handle test cases that involve wet-dry interface, we apply a nondestructive limiter for the fluid depth, and a velocity-based limiter for the momentum, which enables wetting and drying treatment as proposed by Vater et al. \cite{vater2019}. 

To elaborate on each step of the prediction and correction steps, this section is divided into the respective steps. Having the projection method, we can build a locally adaptive model by defining the adaptivity criterion, which we will also discuss briefly in this section.

\subsection{Predictor Step}
We solve our predictor, which is the hydrostatic SWE, with a second-order RK-DG scheme, with a nondestructive limiter for the fluid depth and a velocity-based limiter for the momentum as proposed by Vater et al. \cite{vater2019}. In the compact form, the hydrostatic part can be written as
\begin{equation}\label{eq:genericbalance}
    \tilde{\boldsymbol{Q}}_t+\nabla\cdot\boldsymbol{F}(\tilde{\boldsymbol{Q}}) = \boldsymbol{S}(\tilde{\boldsymbol{Q}}),
\end{equation}
with $\tilde{\boldsymbol{Q}} = (h, h\boldsymbol{u}, hw)^T$ is the unknowns and $\boldsymbol{F}(\tilde{\boldsymbol{Q}}) = \left(h\boldsymbol{u},h\boldsymbol{u}\otimes\boldsymbol{u}+\frac{g}{2}h^2\boldsymbol{I}_2,h\boldsymbol{u}w\right)^T$ and $\boldsymbol{S}(\tilde{\boldsymbol{Q}}) = (0, gh\nabla d, 0)^T$ are the flux and the source terms respectively. 

We discretize our spatial domain $\Omega\in\mathbb{R}^2$, with triangular elements $K_i$. For the weak DG formulation, we multiply equation \eqref{eq:genericbalance} by a test function $\varphi$, integrate it over an element, and apply integration by parts to obtain
\begin{equation}\label{eq:DG_weak}
    \int_{K_i}\varphi\boldsymbol{\tilde{\boldsymbol{Q}}}_td\boldsymbol{x}-\int_{K_i}\nabla\varphi\cdot\boldsymbol{F}(\tilde{\boldsymbol{Q}})d\boldsymbol{x}+\int_{\partial K_i}\varphi\boldsymbol{n}\cdot\boldsymbol{F}^*(\tilde{\boldsymbol{Q}})d\boldsymbol{x}=\int_{K_i}\varphi\boldsymbol{S}(\tilde{\boldsymbol{Q}})d\boldsymbol{x},
\end{equation}
where $\boldsymbol{n}$ represents the outward pointing normal vector on the edges of $K_i$. The communications between adjacent elements are controlled through the interface flux $\boldsymbol{F}^*$, which in this case is defined by a Riemann solver, namely the Rusanov solver \cite{RUSANOV1962304}. We approximate the solution and test function with a linear function built with nodal Lagrange basis functions \cite{giraldo2008, HesthavenWarburton2008} as $\boldsymbol{Q}|_{K_i}\approx\boldsymbol{Q}_h|_{K_i}(\boldsymbol{x},t) = \sum_j(\tilde{\boldsymbol{Q}}_h|_{K_i}(t))_j\varphi_j(\boldsymbol{x})$, with $\tilde{\boldsymbol{Q}}_h|_{K_i}(t)$ is the degree of freedom vector. Moreover, the flux and source terms are approximated similarly. This spatial discretization leads us to the remaining semi-discrete system of ordinary differential equations 
\begin{equation}\label{eq:swe_ode}
    \frac{d\tilde{\boldsymbol{Q}}_h}{dt} = \mathcal{H}(\tilde{\boldsymbol{Q}}_h),
\end{equation}
where $\mathcal{H}$ represents the discretized flux and source terms.

For the time stepping, we discretize \eqref{eq:swe_ode} with an explicit second-order Runge-Kutta time integration, also known as Heun's method. This leads to the following scheme: 
\begin{equation}\label{eq:heun_step2}
    \tilde{\boldsymbol{Q}}_h^{n+1} = \tilde{\boldsymbol{Q}}_h^{n} + \frac{\Delta t}{2}\left(\mathcal{H}(\tilde{\boldsymbol{Q}}_h^n) + \mathcal{H}(\tilde{\boldsymbol{Q}}_h^{\star})\right),
\end{equation} 
with
\begin{equation}\label{eq:heun_step1}
    \tilde{\boldsymbol{Q}}_h^{\star} = \tilde{\boldsymbol{Q}}_h^{n} + \Delta t\left(\mathcal{H}(\tilde{\boldsymbol{Q}}_h^n)\right).
\end{equation}
We call \eqref{eq:heun_step1} and \eqref{eq:heun_step2} the first and second RK stages. However, solving our corrector with the implicit Euler method, which is only first-order accurate, will cost us an order of accuracy if we apply the correction at each time step, as observed by Schlottke-Lakemper et al. \cite{SCHLOTTKELAKEMPER2021110467}. Instead, we correct once every RK stage to preserve the order of accuracy, formulated with 
\begin{equation}\label{eq:heun_step2_2ndorder}
    \tilde{\boldsymbol{Q}}_h^{n+1} = \tilde{\boldsymbol{Q}}_h^{n} + \frac{\Delta t}{2}\left(\mathcal{H}(\tilde{\boldsymbol{Q}}_h^n) + \mathcal{H}(\boldsymbol{Q}_h^{\star})\right),
\end{equation} 
where $\boldsymbol{Q}_h^{\star}$ is a corrected predictor \eqref{eq:heun_step1}.

\subsection{Corrector Step}
Following the predictor step, we solve the remaining terms based on the computed predictor solution with the implicit Euler's method. Note that there are no non-hydrostatic terms involved in the mass conservation equation \eqref{eq:swe_mass}. Hence, we can adapt the predictor as our final solution, i.e., $h^{n+1} = \tilde{h}^{n+1}$. The unknown momentum, on the other hand, can be obtained by solving \eqref{eq:swe_momentum_horizontal} and \eqref{eq:swe_momentum_vertical} with 
\begin{equation}\label{eq:correction_horizontal}
    \frac{(h\boldsymbol{u})^{}-(\tilde{h\boldsymbol{u}})}{\Delta t} = -\frac{1}{\rho}\nabla(\tilde{h}p^{nh})+\frac{1}{\rho}\left(P^{nh}|_{z=-d}(p^{nh},\tilde{\textbf{Q}})\right)\nabla d,
\end{equation}
\begin{equation}\label{eq:correction_vertical}
    \frac{(hw)-(\tilde{hw})}{\Delta t} = \frac{1}{\rho}\left(P^{nh}|_{z=-d}(p^{nh},\tilde{\textbf{Q}})\right).
\end{equation}
For clarity, we neglect the time-step superscript and the discrete-approximation subscript. To solve the last two equations, we need to involve the divergence constraint \eqref{eq:swe_constraint} to close the system. Substituting $h\boldsymbol{u}$ and $hw$ from \eqref{eq:correction_horizontal} and \eqref{eq:correction_vertical} to \eqref{eq:swe_constraint}, yields an elliptic equation for the unkown $p^{nh}$. We solve this elliptic problem with the local discontinuous Galerkin (LDG) method, which decomposes the second-order operator into a series of two first-order operators \cite{giraldo2020}. We may think of $h\boldsymbol{u}$ and $hw$ as the auxiliary variables constructed through \eqref{eq:correction_horizontal} and \eqref{eq:correction_vertical}.

Discretizing \eqref{eq:correction_horizontal} and \eqref{eq:correction_vertical} spatially with the discontinuous Galerkin method, where we approximate $h\boldsymbol{u},~hw,$ and $p^{nh}$ with a piecewise linear function, we can write
\begin{equation}\label{eq:disc_momentum}
    q = L_{p^{nh}}^{(q)}p^{nh}+S^{(q)},
\end{equation}
with $q$ being either horizontal ($q=hu$ or $hv$) or vertical ($q=hw$) momentum. The discretized coefficients of $p^{nh}$ and constants are represented by $L_{p^{nh}}^{(q)}$ and $S^{(q)}$ respectively. Similarly, the divergence constraint \eqref{eq:swe_constraint} is discretized as
\begin{equation}\label{eq:disc_constraint}
    L_{(hu)}(hu)+L_{(hv)}(hv)+L_{(hw)}(hw) = S,
\end{equation}
with $L$ and $S$ being discretized coefficients and constants, respectively. To solve the elliptic problem, we first evaluate \eqref{eq:disc_momentum}, which is then used to construct \eqref{eq:disc_constraint}. This leads to a system of linear equations for $p^{nh}$, which can be solved with an iterative method, such as the Biconjugate Gradients Stabilized method. 

For terms that involve the unknown gradient, we need to define the numerical flux. We simply take a central flux, defined as 
\begin{equation}
    (\boldsymbol{q})^{*} = (\boldsymbol{q}^{+}+\boldsymbol{q}^{-})/2,
\end{equation}
\begin{equation}
    (p^{nh})^{*} = ((p^{nh})^++(p^{nh})^-)/2,
\end{equation}
where the superscript "$+$" and "$-$" denotes the value on two adjacent elements $K_+$ and $K_-$ respectively.

\subsection{Locally Adaptive Model}
By solving our model with a projection method, we may adapt the correction locally in regions where non-hydrostatic pressure might be crucial. In one-dimensional settings, it has been shown that the locally adaptive model works well with various test cases, ranging from a solitary wave to moving bottom-generated waves \cite{firdaus2026}. Such a model requires a proper criterion, which relies on the calculated predictor values, namely the hydrostatic SWE solution.

\begin{figure}[h!]
    \centering
    \subfigure{%
        \includegraphics[width=0.3\textwidth]{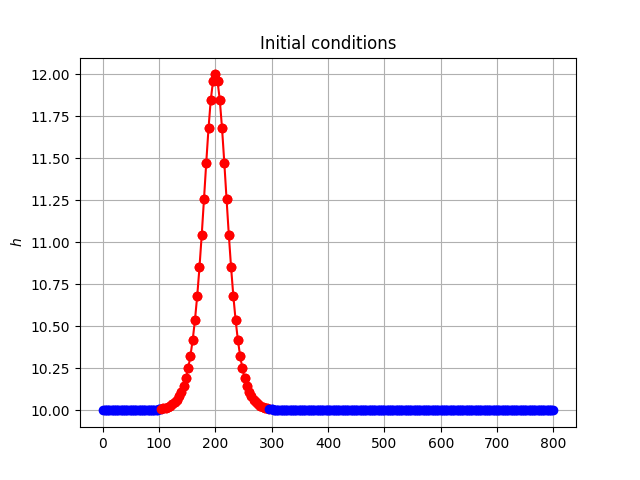}
    }
    \subfigure{%
        \includegraphics[width=0.3\textwidth]{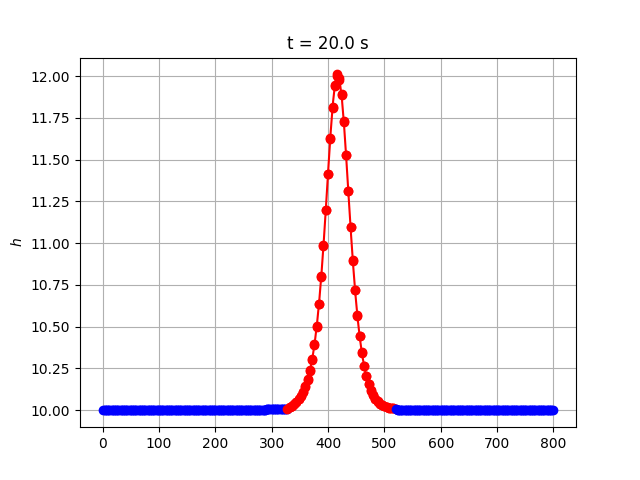}
    }
    \subfigure{%
        \includegraphics[width=0.3\textwidth]{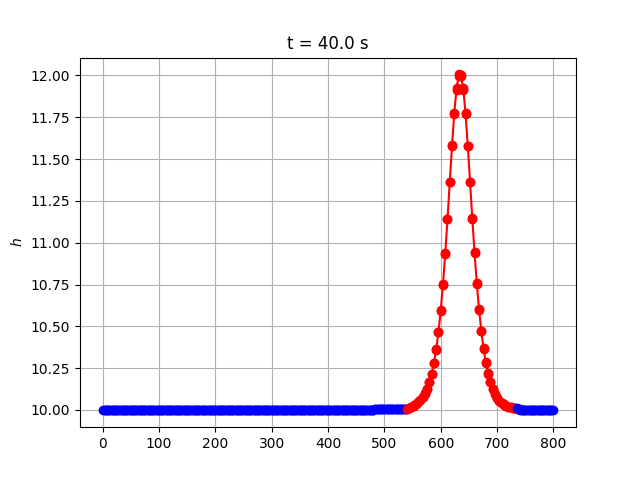}
    }
    \caption{Locally adaptive simulation of propagating solitary wave, where the corrections are adapted when $|\tilde{\eta}/d|>0.001$, at time stamps $t=0, 20,$ and $40~s$.}
    \label{fig:result_solitary}
\end{figure}

\begin{figure}[h!]
    \centering
    \subfigure{%
        \includegraphics[width=0.3\textwidth]{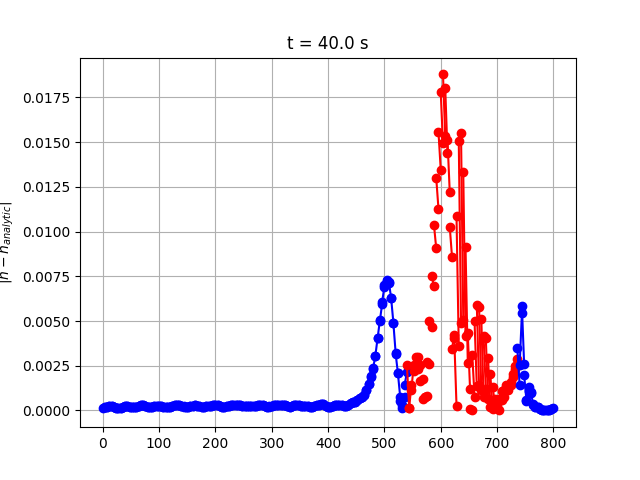}
    }
    \subfigure{%
        \includegraphics[width=0.3\textwidth]{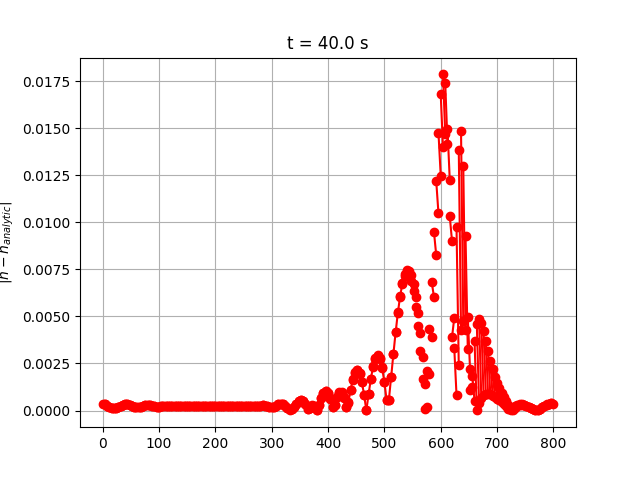}
    }
    \caption{Absolute error produced by locally adaptive (left) and global (right) models of the surface elevation of the solitary wave simulation.}
    \label{fig:error_solitary}
\end{figure}

To give a brief illustration, Figure \ref{fig:result_solitary} shows the simulated solitary wave with a locally adaptive model, where the corrected domain is marked with the red area. This corrected domain is based on the magnitude of the elevation and depth ratio $|\tilde{\eta}/d|>0.001$. At the end of the simulation ($t=40~s$), both the global and local models produce a similar absolute error (see Figure \ref{fig:error_solitary}), with the latter approach saving almost $75\%$ of the computational time. In this study, we use the combination of the magnitude of the elevation-depth ratio ($|\tilde{\eta}/d|>0.001$) with the horizontal-velocity norm ($\lVert\tilde{\boldsymbol{u}}\rVert_2>0.001$).

\section{Results and Discussion}
To assess the validity of our model, we apply it to three laboratory-based test cases. We begin with a test case that involves wave shoaling over a sloping bottom, while the two following cases imitate tsunami wave propagation, involving both a static and a moving bottom. 

In each case, we examine two objectives. First, we want to know how suitable the model is for each case, more precisely, how the results obtained from each linear and quadratic pressure assumption are. This is reflected through the comparison of the global model with the measured data. Moreover, we aim to determine how close the results obtained from the local model align with those from the global model and how much computation time can be reduced. Hence, we compare the results from the global and local models for each linear and quadratic pressure relation, as well as the corrected elements and computational time ratio.

\subsection{Periodic waves propagation over a semi-circular shoal}
\begin{figure}[h]
    \centering
    \includegraphics[width=.6\linewidth]{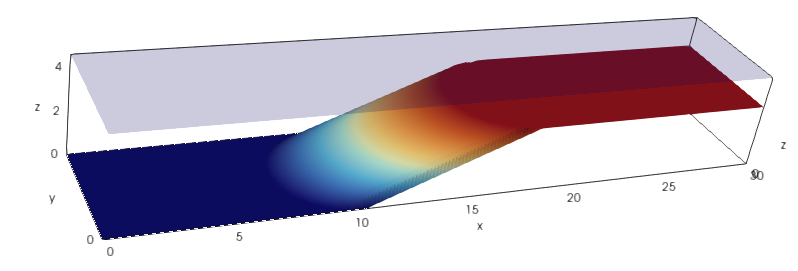}
    \caption{Three-dimensional view of the semi-circular shoal rescaled by a factor of ten in the vertical direction.}
    \label{fig:bathy_whalin}
\end{figure}

Our first benchmark is based on an experiment by Whalin\cite{Whalin1971} that assessed the limits of applicability of linear wave refraction theory in a convergence zone. The experiment was performed in a $25.6~m\times6.096~m$ basin, where a semi-circular shoal was installed in its center portion, yielding an initial condition of decreasing undisturbed water with a depth of $0.4572~m$ to $0.1524~m$ (see Figure \ref{fig:bathy_whalin}). This bathymetry can be expressed as
\begin{equation}
    d(x,y,t) = \begin{cases}
        0.4572&,\ x\leq10.67-\Gamma(y),\\
        0.4572+\frac{10.67-\Gamma(y)-x}{25}&,\ 10.67-\Gamma(y)\leq x\leq 18.29-\Gamma(y),\\
        0.1524&,\ x>18.29-\Gamma(y),
    \end{cases}
\end{equation}
with $\Gamma(y) = \sqrt{y(6.096-y)}$. Wave trains are generated from the left-hand side of the domain ($x=0$).

Our computational domain is extended by $(0,30)\times(0,6.096)~m^2$ to define a  zone of linearly-increasing quadratic bottom friction for $x>25.6~m$, avoiding any reflected waves. We discretize it by $196\times32$ uniform rectangular cells, which are then divided into four triangles, yielding $24576$ triangular cells. The wave trains are generated from the left side of the boundary, which is then absorbed downstream, while wall boundary conditions are imposed on the lateral boundaries. We consider two test cases involving wave periods of $T = 2$ and $3~s$ corresponding to an amplitude of $a = 0.0075$ and $0.0068~m$. The simulations run for $100~s$, with a time step of $\Delta t=0.01~s$.

\begin{figure}[h]
    \centering
    \includegraphics[width=.75\linewidth]{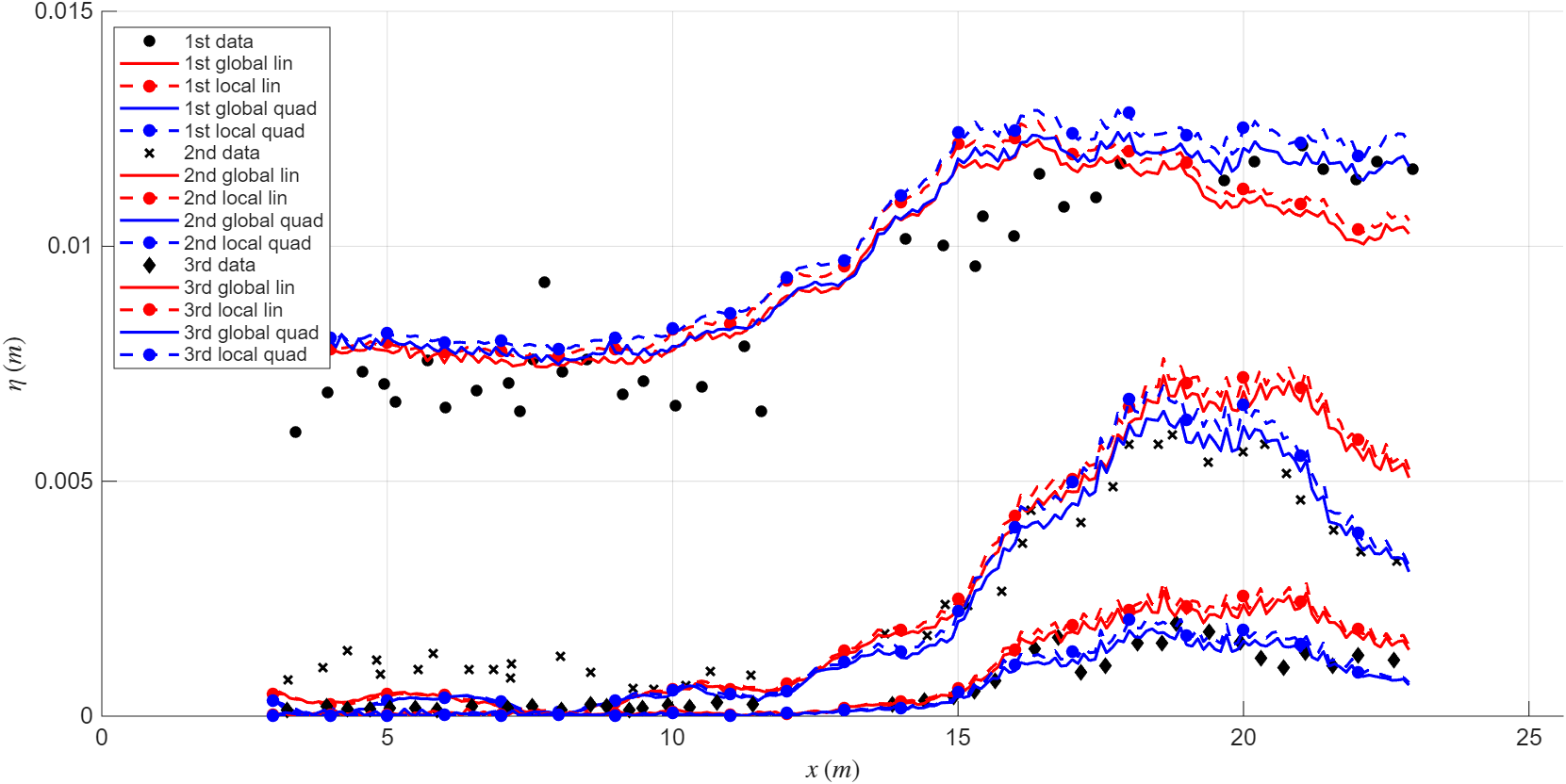}
    \caption{Comparison of the measured data (black solid line) with the global (colored solid line) and the locally adaptive (colored dashed line) simulations for the first, second, and third harmonics for the case $T=2~s$. Simulations are done using linear (red) and quadratic (blue) pressure relations.}
    \label{fig:result_whalin_T2}
\end{figure}

\begin{figure}[h]
    \centering
    \includegraphics[width=.75\linewidth]{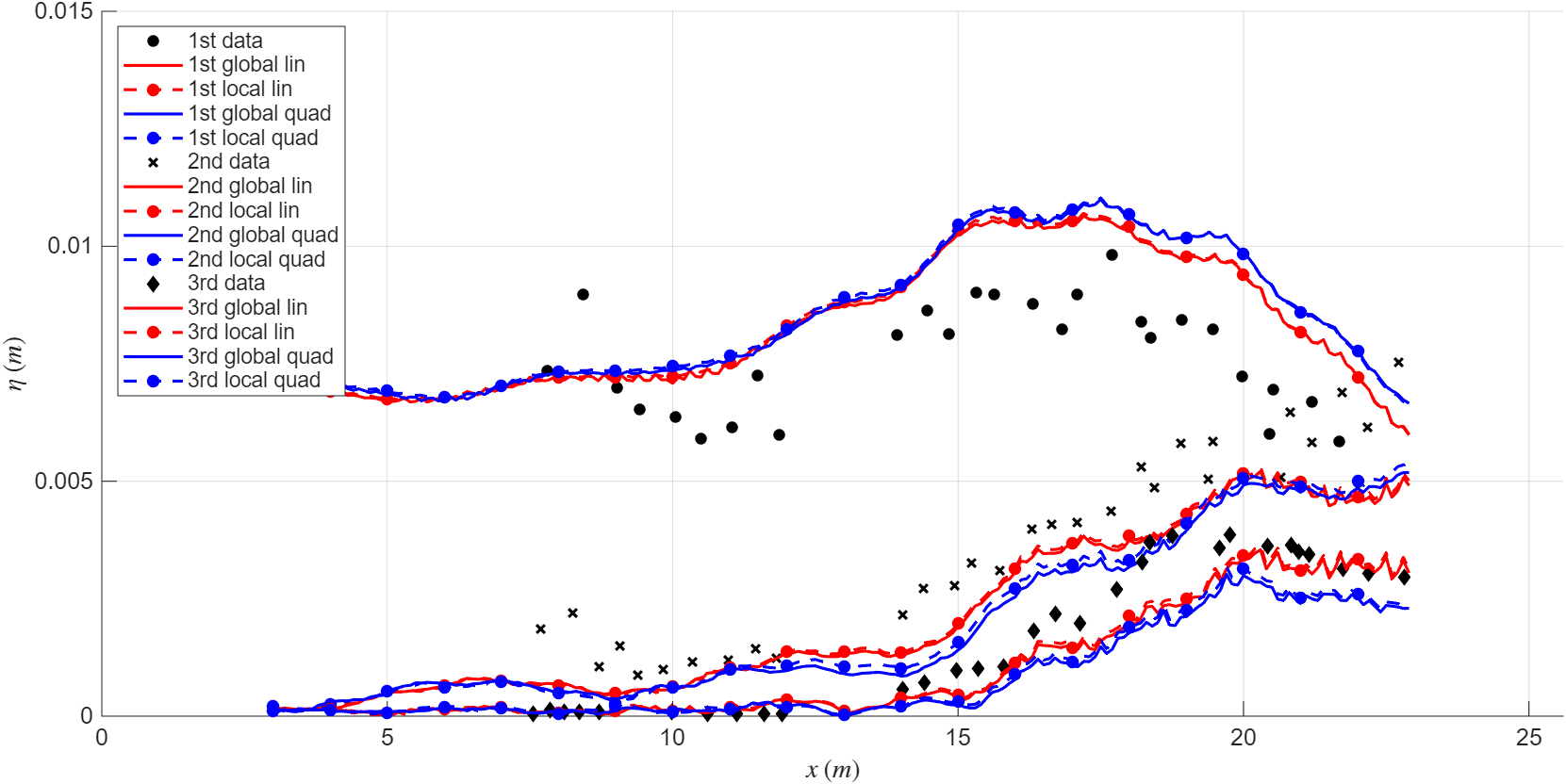}
    \caption{Comparison of the measured data (black solid line) with the global (colored solid line) and the locally adaptive (colored dashed line) simulations for the first, second, and third harmonics for the case $T=3~s$. Simulations are done using linear (red) and quadratic (blue) pressure relations.}
    \label{fig:result_whalin_T3}
\end{figure}

We analyze the surface elevation along the centerline ($y=3.048$) for the last $25~s$ of simulation time. A harmonic analysis is conducted to obtain the first, second, and third harmonic amplitudes. Figure \ref{fig:result_whalin_T2} and \ref{fig:result_whalin_T3} compare the harmonic amplitudes with the measured data for $T=2$ and $3~s$ respectively. For the case with $T=2~s$, the simulation produced with the quadratic pressure gives more accurate results for the second and third harmonics. In contrast, the linear pressure tends to overestimate the amplitude of the second and third harmonics after passing the shoal. In case $T=3~s$, notable disagreement can be observed for both pressure relations, where our results overshoot the first harmonic and undershoot the second and third harmonics. These discrepancies are also observed in previous studies involving various Boussinesq-type equations, e.g. \cite{MADSEN1992183, KAZOLEA201242, LANNES2015238}.

\begin{figure}[h!]
    \centering
    \subfigure{%
        \includegraphics[width=0.45\textwidth]{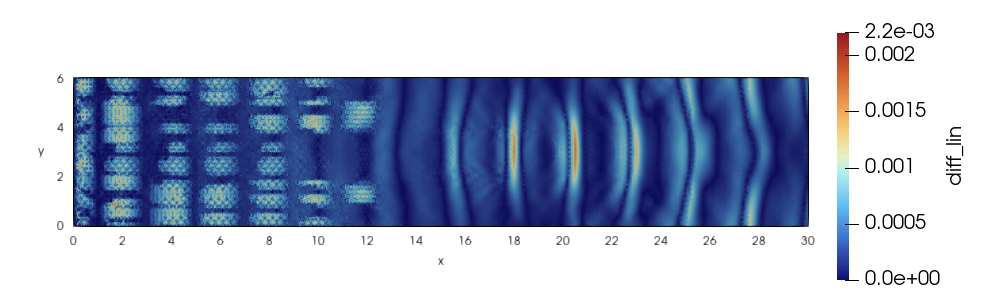}
        \label{fig:diff_lin_whalin_T2}
    }
    \subfigure{%
        \includegraphics[width=0.45\textwidth]{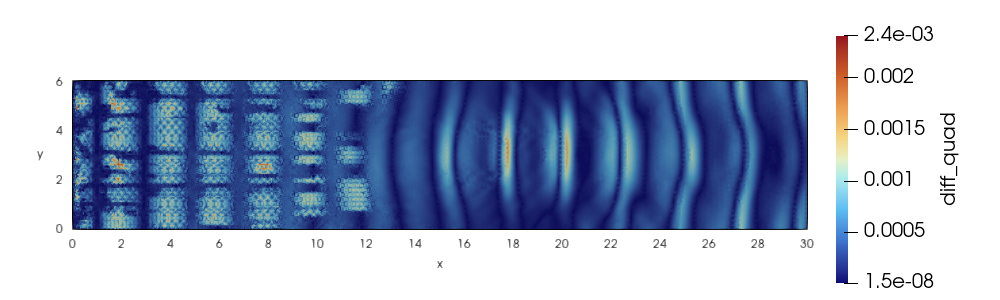}
        \label{fig:diff_quad_whalin_T2}
    }
    \subfigure{%
        \includegraphics[width=0.45\textwidth]{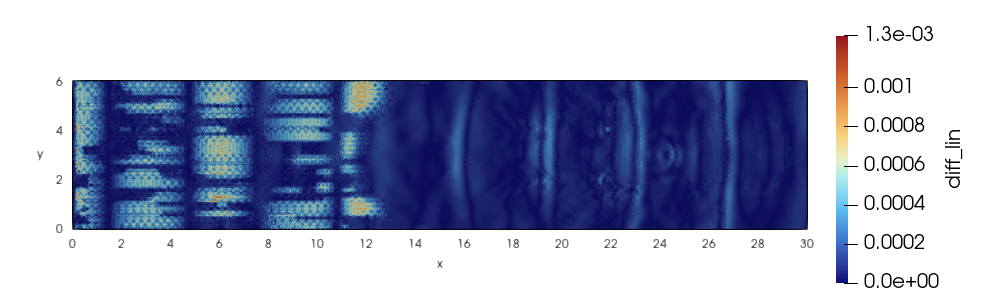}
        \label{fig:diff_lin_whalin_T3}
    }
    \subfigure{%
        \includegraphics[width=0.45\textwidth]{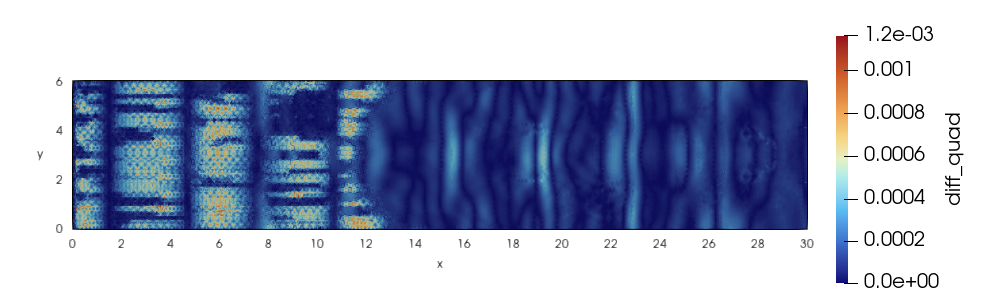}
        \label{fig:diff_quad_whalin_T3}
    }
    \caption{Absolute difference of the simulated elevation by locally adaptive and global models at the end of the simulation time from linear (left) and quadratic (right) pressure relations, for $T=2$ (top) and $3~s$ (bottom), respectively.}
    \label{fig:diff_whalin}
\end{figure}

\begin{figure}[h!]
    \centering
    \subfigure{%
        \includegraphics[width=0.6\textwidth]{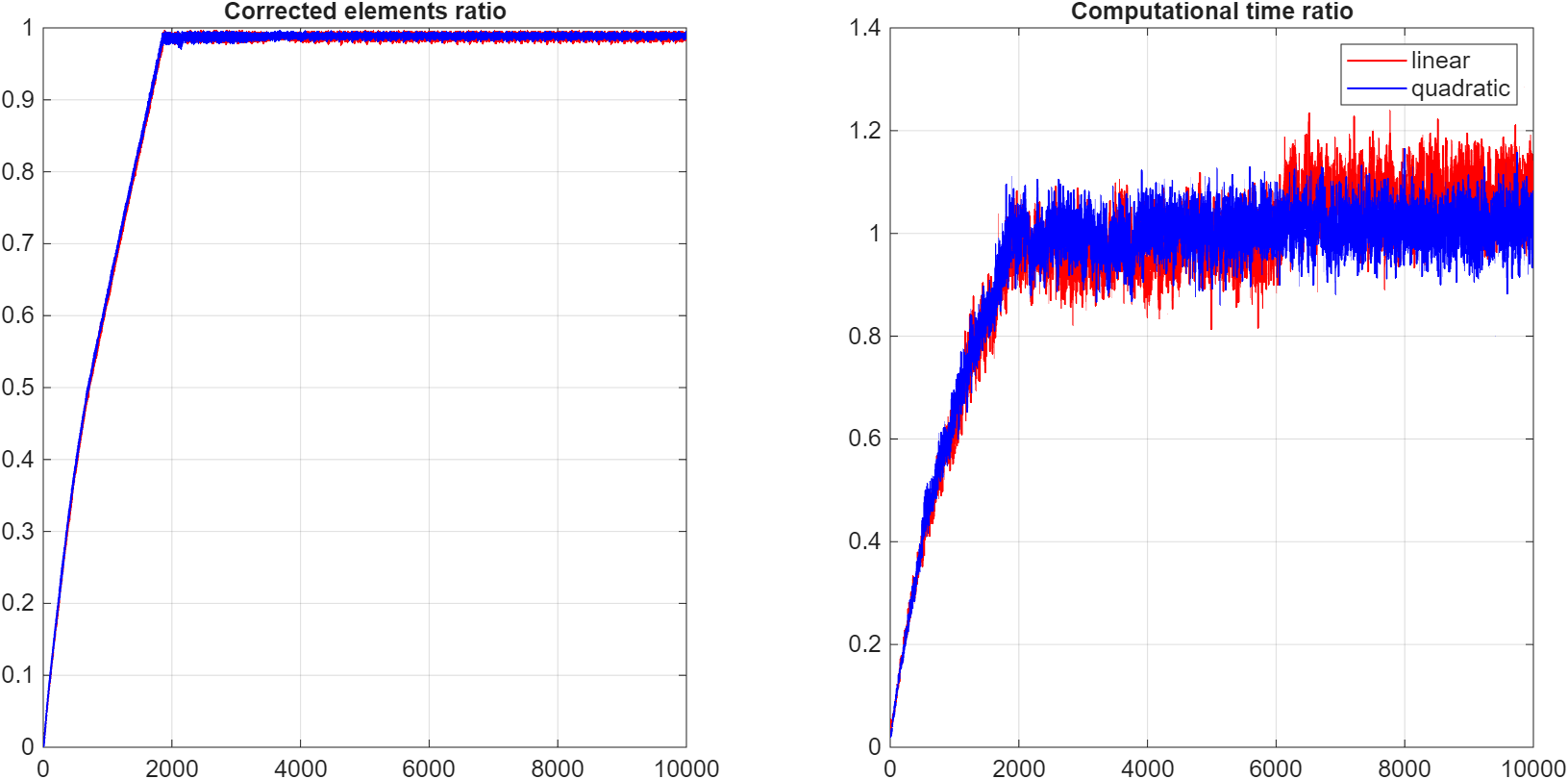}
        \label{fig:ratio_whalin_T2}
    }
    \vspace{0.5cm}
    \subfigure{%
        \includegraphics[width=0.6\textwidth]{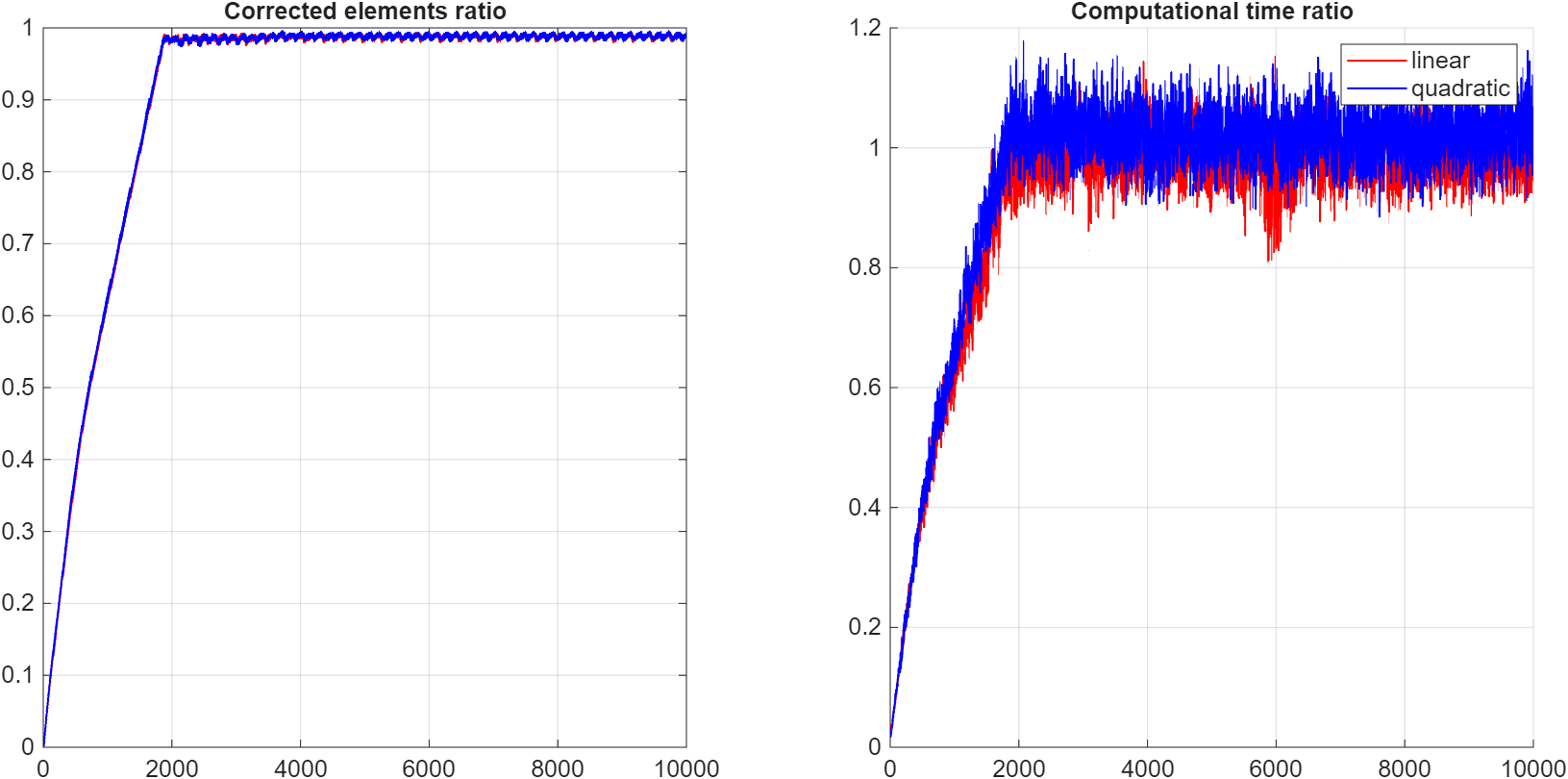}
        \label{fig:ratio_whalin_T3}
    }
    \caption{Ratio of corrected elements (left) and computational time (right) across time steps for the locally adaptive model versus the global model, for $T=2$ (top) and $3~s$ (bottom).}
    \label{fig:ratio_whalin}
\end{figure}

The local model generally yields results close to those of the global model, as shown by the absolute difference at the end of the simulation in Figure \ref{fig:diff_whalin}. The absolute difference is bounded by less than two orders of magnitude smaller than the fluid depth.
Figure \ref{fig:ratio_whalin} depicts the ratio of the number of corrected elements to the total number of elements, along with the computational time over time steps.  Since this case involves generated wave trains, we can observe that the number of corrected elements ratio increases over time until it approaches one and remains steady. This is due to the fact that non-hydrostatic wave dispersion eventually covers the whole domain. The same applies to the computational time, yet there is overhead when the correction is applied to most of the domain, because of the cost for computing the criterion and the dynamic adaptation of the system matrix size as the number of corrections varies. Nevertheless, the local model still requires less computational time, accounting for $93.40\%$ and $92.78\%$ of the total time for the $T=2~s$ case, and $90.63\%$ and $94.22\%$ for the $T=3~s$ case, when using linear and quadratic relations, respectively. This situation occurs in the present wave train simulation, where the scattered waves propagate throughout the entire computational domain, causing the correction to be activated over a large region and thereby reducing the potential computational benefits of the local approach. The proposed local model remains particularly advantageous for scenarios involving a limited region of wave activity, which is encountered in tsunami waves.

\subsection{Flow over a conical island}
This test is based on an experiment conducted by the U.S. Army Engineer Waterways Experiment Station, which was motivated by the tsunami runup on Babi Island in 1992 \cite{Briggs1995, Liu_Cho_Briggs_Kanoglu_Synolakis_1995}. The basin dimension was $25~m\times28.2~m$, where a conical island was located in the center to simplify the island shape. This basin was filled with water to a height of $0.32~m$. A solitary wave was generated from one side of the domain. Several solitary wave setups were considered, from which we use the steepest one, denoted as case C, involving an amplitude of $\eta_0 =0.057~m$.

\begin{figure}[h]
    \centering
    \includegraphics[width=.5\linewidth]{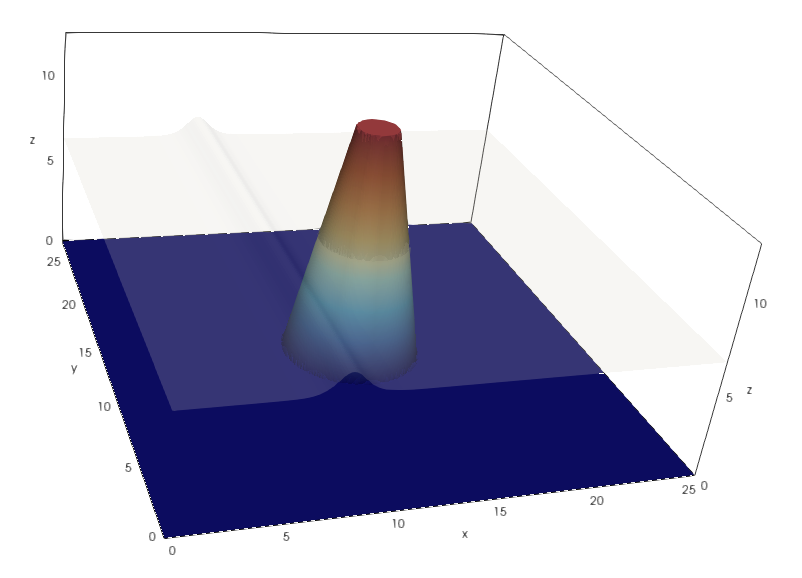}
    \caption{Three-dimensional view of the conical island rescaled by a factor of twenty in the vertical direction.}
    \label{fig:bathy_conical}
\end{figure}

We adjust our domain to $[0~m,25.92~m]\times[0~m,27.60~m]$, and discretize it by $131072$ triangular elements, consisting of $256\times256$ squares divided into two triangles. Transparent boundary conditions are employed on the left and right sides of the domain, while hard wall boundary conditions are imposed on the lateral boundaries. Instead of generating the wave from the boundary, we define the solitary wave as an initial condition defined as
\begin{equation}
    h(x,y,0) = \max\left\{0,d_0+\frac{\eta_0}{\cosh^2{(K(x-x_0))}}\right\},
\end{equation}
which represents a solitary wave along $x$-axis with an amplitude of $\eta_0$ centered along $x_0$ with a scaling factor $K=\sqrt{0.75\eta_0/d_0^3}$, where $d_0=0.32~m$ represents the water depth at rest (see Fig. \ref{fig:bathy_conical} for illustration). We chose $x_0 = 7.56~m$ such that the solitary wave height drops to $5\%$ at the toe of the island. Physically, it represents propagating solitary waves with the maximum amplitude located along $x=x_0$, yielding a time shift of $7.77~s$ in our simulation. Due to the model time shift, we run our simulation until $t=12.23~s$, with a time step of $\Delta t=0.01$.

\begin{figure}[h]
    \centering
    \includegraphics[width=1\linewidth]{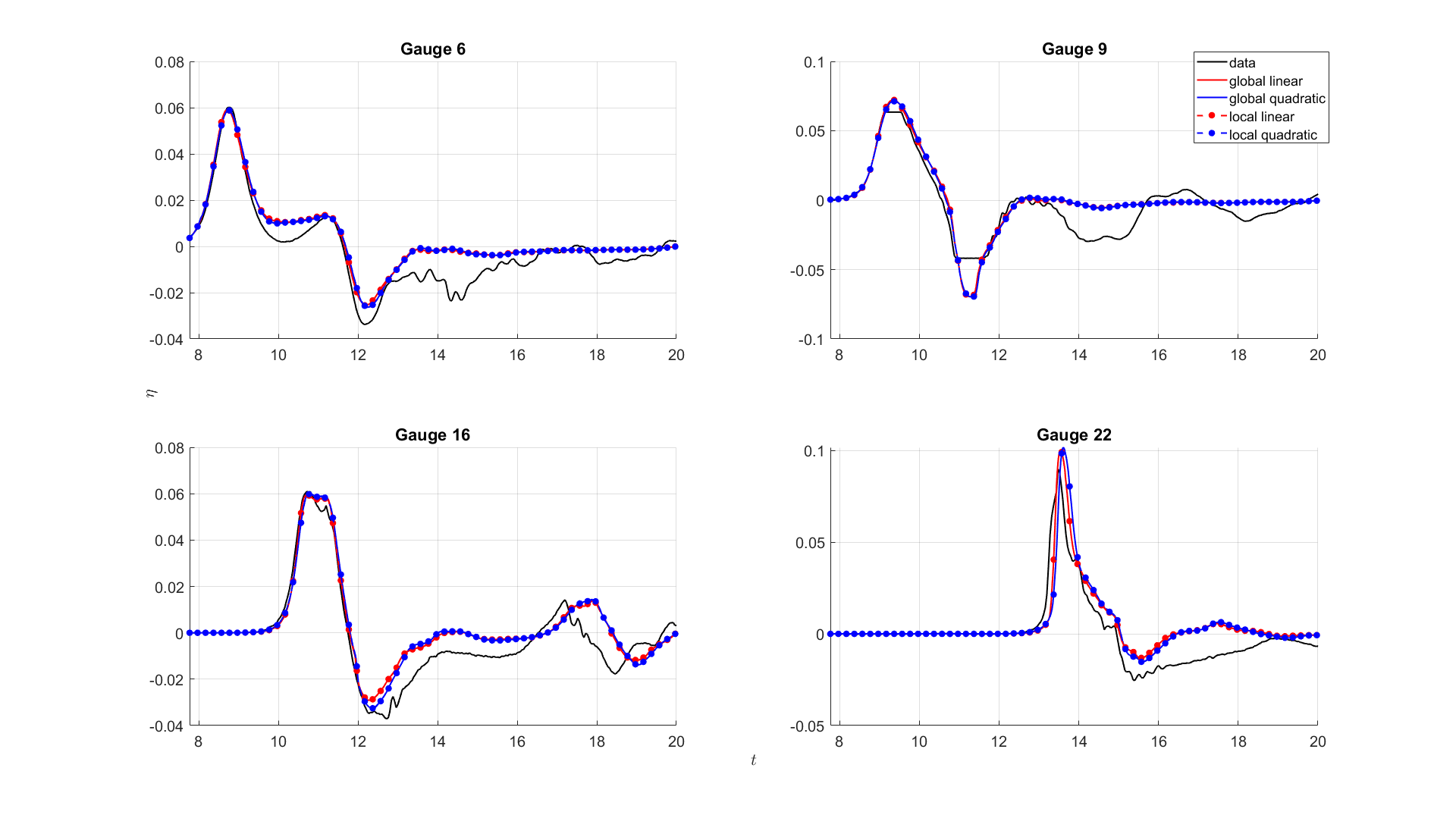}
    \caption{Comparison of the measured data (black solid line) with the global (colored solid line) and the locally adaptive (colored dashed line) simulations for the surface elevation recorded at wave gauges $6, 9, 16, 22$ (top left to bottom right). Simulations are done using linear (red) and quadratic (blue) pressure relations.}
    \label{fig:result_conical_C}
\end{figure}

Figure \ref{fig:result_conical_C} shows the comparison of our model with both linear and quadratic pressure profiles, along with the locally adaptive model against the laboratory-measured data, which is shifted for $20~s$. We compare the elevation at four stations: $6, 9, 16$ and $22$, which correspond to $(x,y) = (9.36,13.8), (10.36,13.8), (12.96,11.22)$, and $(15.56,13.8)$ respectively. These stations cover the front, side, and back sides of the island. 
In general, both results from linear and quadratic pressure relations agree well with the data. Both relations behave similarly in this case, especially at the front side (gauges $6$ and $9$). A slightly higher estimate is observed for the linear relation as the wave moves downward along the sides (gauge $16$) and the rear of the island (gauge $22$).

\begin{figure}[h!]
    \centering
    \subfigure{%
        \includegraphics[width=0.35\textwidth]{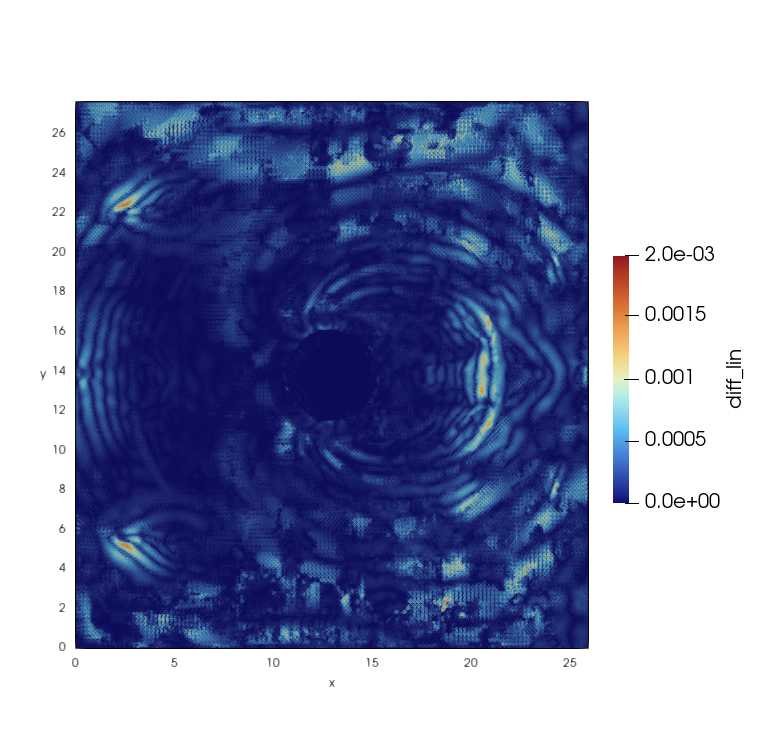}
    }
    \subfigure{%
        \includegraphics[width=0.35\textwidth]{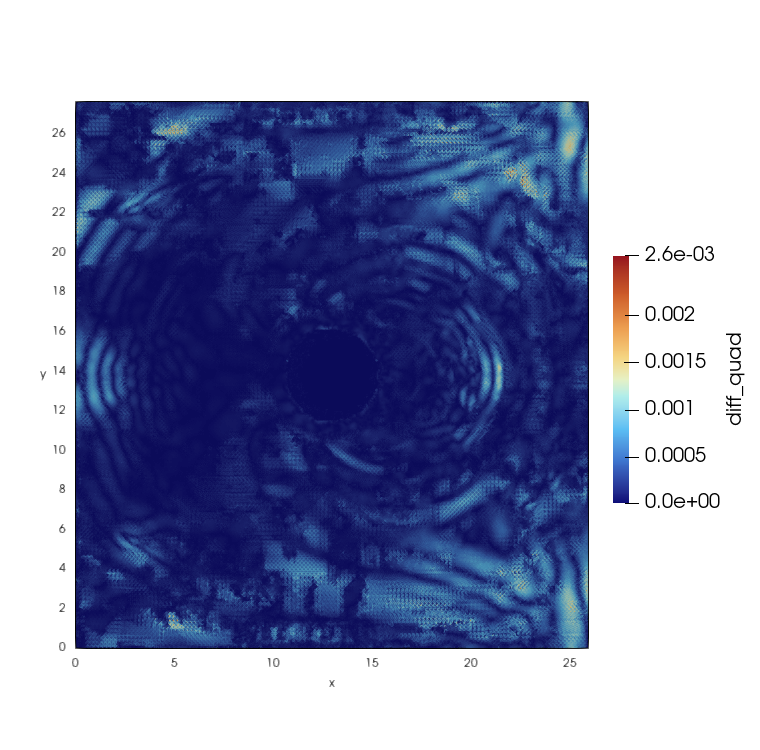}
    }
    \caption{Absolute difference of the simulated elevation by locally adaptive and global models at the end of the simulation time. Simulations are done with linear (left) and quadratic (right) pressure relations.}
    \label{fig:diff_conical_C}
\end{figure}

\begin{figure}[h!]
    \centering
    \includegraphics[width=.6\linewidth]{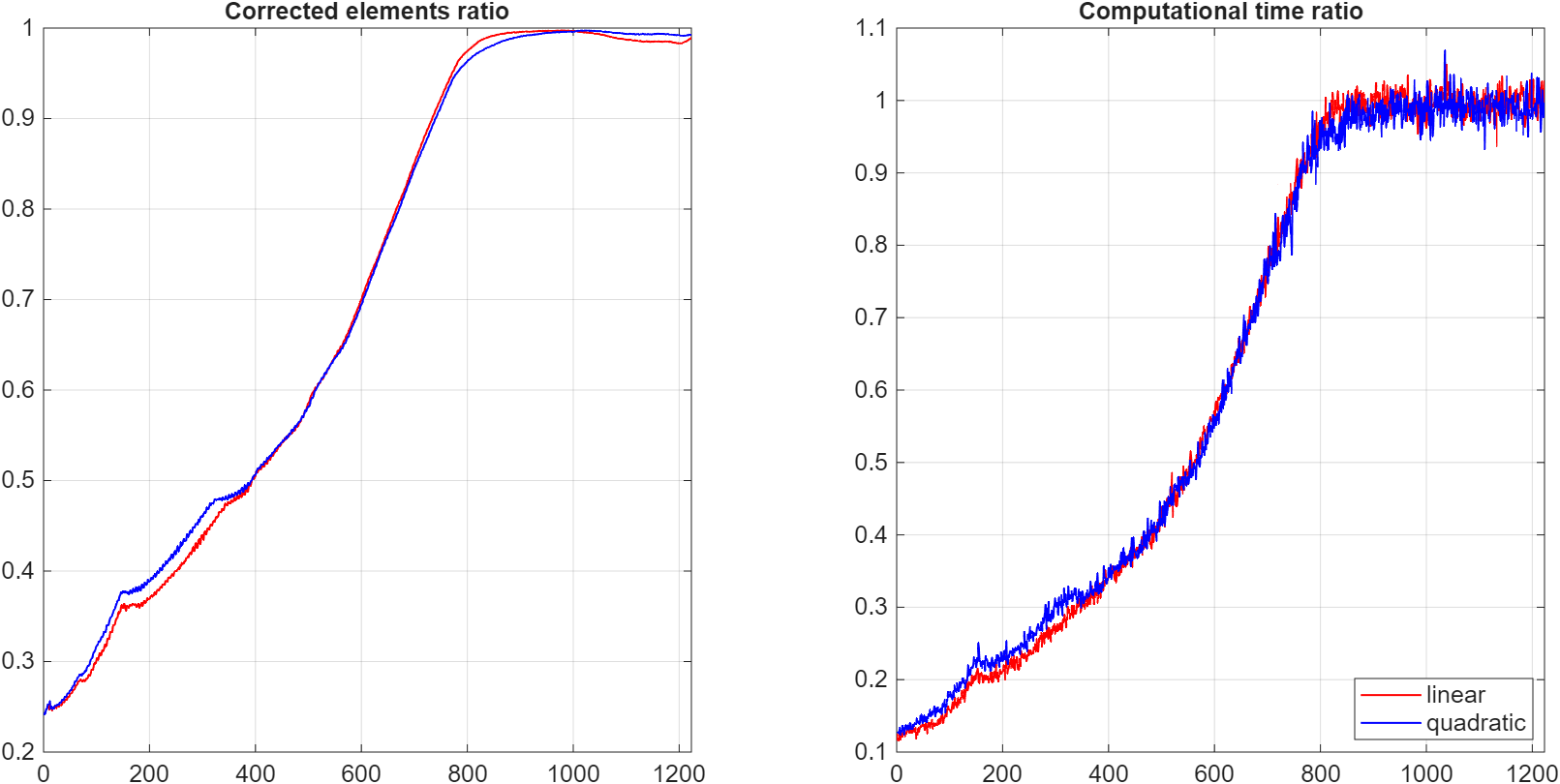}
    \caption{Ratio of corrected elements (left) and computational time (right) of the locally adaptive model compared to the global model over time steps.}
    \label{fig:ratio_conical_C}
\end{figure}

The local model, on the other hand, gives a very close result to the global model. Figure \ref{fig:diff_conical_C} illustrates the difference between the global and local surface elevations at the end of the simulation time. Some relatively small deviation can be observed, with the maximum pointwise error being limited to two orders of magnitude smaller than the fluid depth. 
The ratio of the number of corrected elements and the computational time evolution can be seen in Figure \ref{fig:ratio_conical_C}. Over the time steps, the corrected elements grow as the wave scatters over the whole domain, reflected by the island.
At the end, it takes $61.44\%$ and $61.7\%$ of the global computational time to compute the local model. This result emphasizes that such phenomena, which imitate a tsunami propagation, could benefit from the proposed locally adaptive model in terms of computational time without losing significant accuracy.

\subsection{Submarine landslide over a sloping bottom}
\begin{figure}[h]
    \centering
    \includegraphics[width=.6\linewidth]{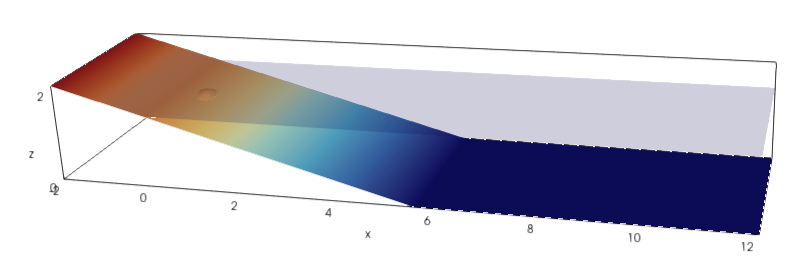}
    \caption{Three-dimensional view of the initial landslide topography.}
    \label{fig:bathy_landslide}
\end{figure}

Motivated by the importance of non-hydrostatic pressure in landslide-generated tsunamis, we apply our model to a moving bottom-generated wave case. This test case is based on a rigid submarine landslide experiment proposed by Enet and Grilli\cite{enet2007}. The experiment was conducted in a water tank measuring $30~m$ in length, $3.7~m$ in width, and $1.8~m$ in depth, filled with $d_0=1.5$ meters of water. The wave was generated by a rigid landslide with a Gaussian hump form with thickness $T = 0.082~m$, length $b = 0.395~m$, and width $w = 0.680~m$ that slides down an incline of $\theta = 15^{\circ}$ (see Figure \ref{fig:bathy_landslide}). Following Fuhrman and Madsen \cite{FUHRMAN2009747}, the time-dependent floor variation can be expressed as
\begin{equation}
    d(x,y,t) = \min\{d_0,-\max\{z_b,-x\tan\theta\}\},
\end{equation}
where $z_b$ is obtained by solving
\begin{equation}
    (\epsilon-1)(z_b\cos\theta+x\sin\theta) = T\left[\epsilon-\cosh^{-1}(k_wy)\cosh^{-1}\left(\frac{k_b}{2}\sec\theta(x-2x_0+x\cos(2\theta)-2S(t)\cos\theta-z_b\sin(2\theta))\right)\right].
\end{equation}
Initially, the mass center base is located at $x_0 = x_g-T\sin\theta$, with $x_g$ being the minimum submergence initial location. The time-dependent function $S(t) = S_0\ln{\left(\cosh\frac{t}{t_0}\right)}$ controls the slide movement, with $S_0=u_t^2/a_0$ and $t_0=u_t/a_0$. We consider a case involving an initial minimum submergence depth $d_g=0.061~m$, corresponding to $x_g=0.551~m$, with an initial landslide acceleration $a_0=1.2~m/s^2$ and terminal landslide velocity $u_t=1.7~m/s$.

 It is worth noting that previous studies that successfully simulated this benchmark used either three-dimensional \cite{AI20211}, multilayer with linear pressure relation \cite{macias2021, TARWIDI2024118750}, or higher-order Boussinesq models \cite{FUHRMAN2009747}. We focus on a one-layer non-hydrostatic SWE extension to assess the performance of our locally adaptive model compared to the global model and to investigate the different behavior of linear versus quadratic pressure relations under a more challenging setup.

Our computational domain is $[-2~m,12~m]\times[-1.85~m,1.85~m]$, which is discretized with $32\times224$ uniform squares that are divided into four triangles, yielding a total of $28672$ triangular elements. We apply an absorbing boundary at the downstream side and hard wall boundary conditions on the rest. The time step is $\Delta t = 0.0035$, which runs until $1000$ time steps. 

\begin{figure}[h!]
    \centering
    \includegraphics[width=1\linewidth]{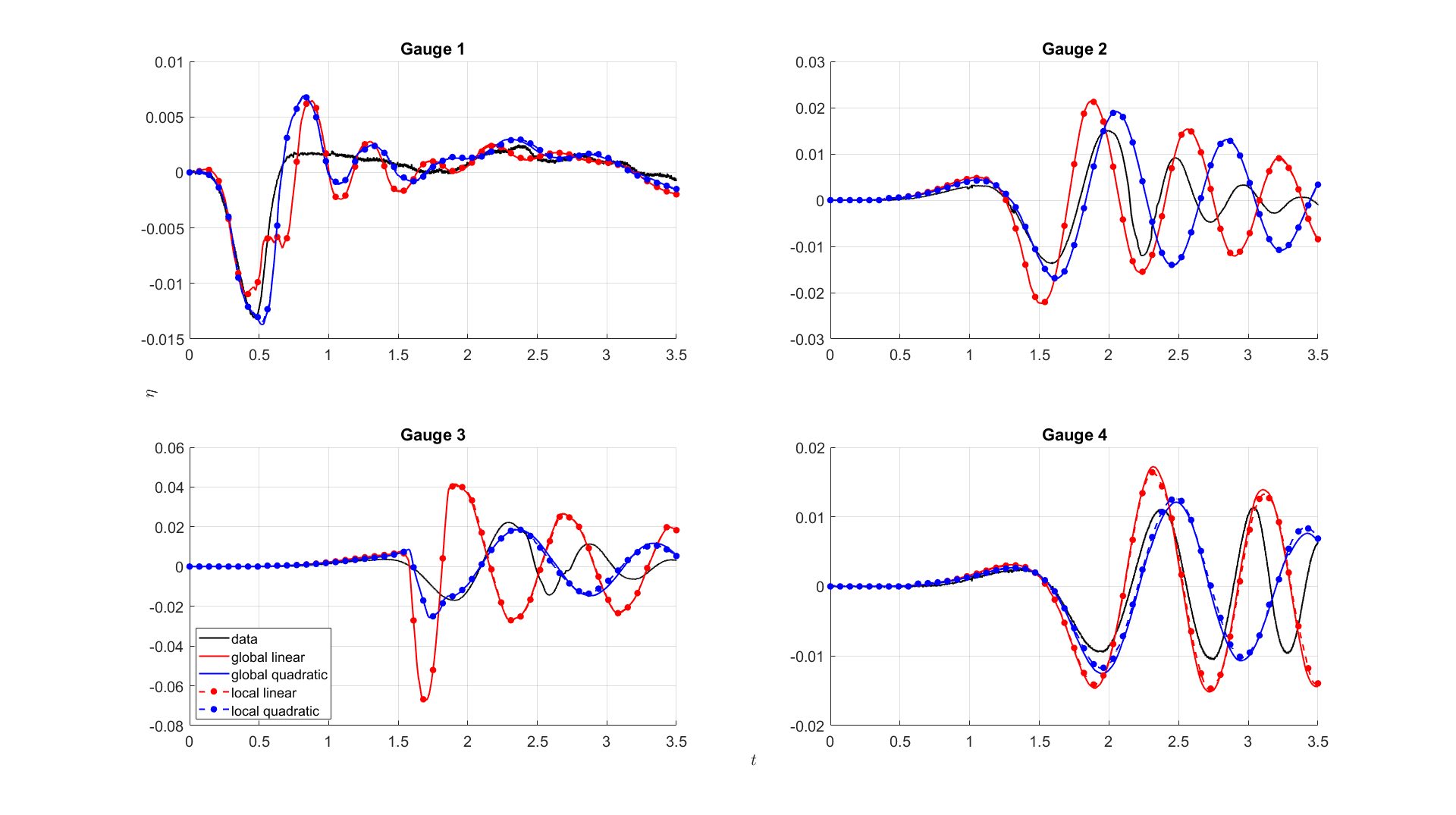}
    \caption{Comparison of the measured data (black solid line) with the global (colored solid line) and the locally adaptive (colored dashed line) simulations for the surface elevation recorded at wave gauges $1,2,3,4$ (top left to bottom right). Simulations are done with linear (red) and quadratic (blue) pressure relations.}
    \label{fig:result_landslide_61}
\end{figure}

From our comparisons in Figure \ref{fig:result_landslide_61}, it is clear that one-layer non-hydrostatic extension of SWE, either with a linear or quadratic pressure relation, is unable to simulate the accurate amplitude or wave period. Different pressure relations give significantly different results. The linear pressure relation tends to overestimate the amplitude, which is also observed by Macías et al. \cite{macias2021} at gauge 4 for the case $d=189~mm$, while the quadratic relation shows a slower wave period. As suggested by Kirby et al. \cite{Kirby2018nthsmp}, most landslide tsunamis are often adequately
described using three vertical layers.

\begin{figure}[h]
    \centering
    \includegraphics[width=.6\linewidth]{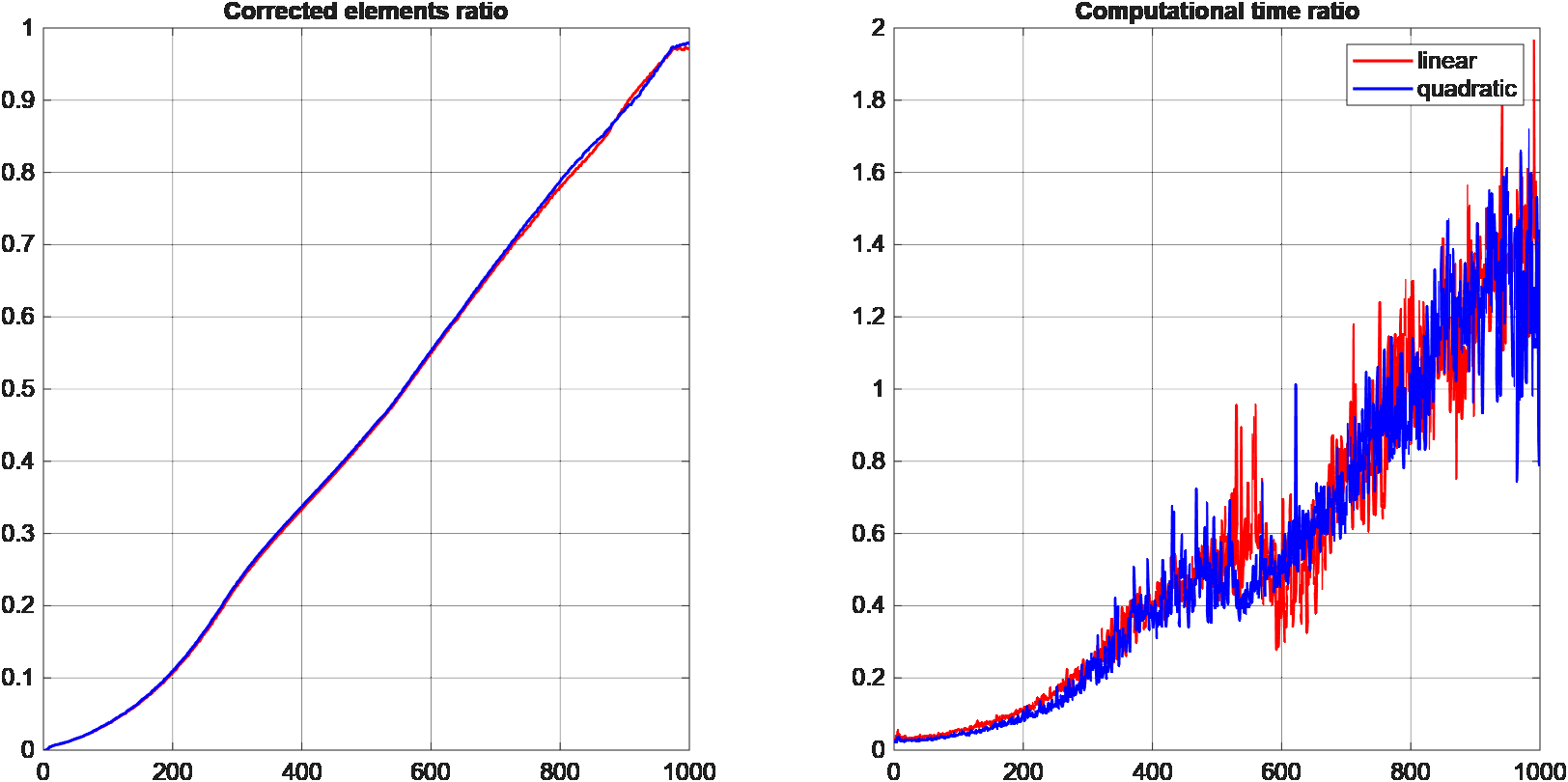}
    \caption{Ratio of corrected elements (left) and computational time (right) of the locally adaptive model compared to the global model with linear (red) and quadratic (right) pressure relations.}
    \label{fig:ratio_landslide_61}
\end{figure}

Despite the limited agreement with experimental data, our locally adaptive approach still manages to give a close result to the global model for such a complex case. The corrected element ratio and the computational time over time steps can be seen in Figure \ref{fig:ratio_landslide_61}. The locally adaptive model takes $65.17\%$ and $61.80\%$ of the computational time for the linear and quadratic pressure, respectively. This case reaffirms that tsunami-type waves can benefit from locally adaptive models.
\begin{figure}[h!]
    \centering
    \subfigure{%
        \includegraphics[width=0.45\textwidth]{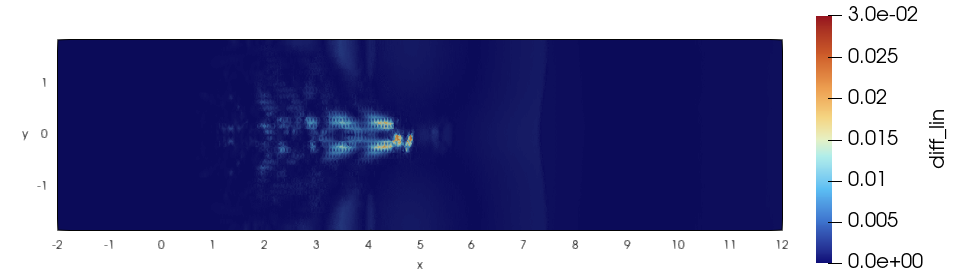}
        \label{fig:diff_lin_landslide_d61}
    }
    \subfigure{%
        \includegraphics[width=0.45\textwidth]{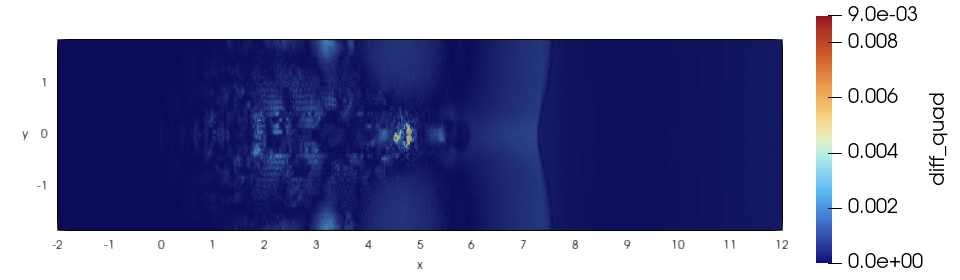}
        \label{fig:diff_quad_landslide_d61}
    }
    \caption{Absolute difference of the simulated surface elevation between locally adaptive and global model with linear (left) and quadratic (right) pressure relation at the end of simulation time.}
    \label{fig:diff_landslide_d61}
\end{figure}

Theoretically, a locally adaptive model can be adapted fairly straightforwardly, provided we have a model that can be solved using a projection method. In fact, some of the previously more advanced models that have successfully simulated this case were solved using a projection method, e.g., Ai et al. \cite{AI20211} with Navier-Stokes equations and  Macías et al., Tarwidi et al. \cite{macias2021, TARWIDI2024118750} with a multilayer model. Hence, further investigation into adapting the locally adaptive approach to more advanced models is of interest.

\section{Conclusions}\label{sec6}
This work introduces a two-dimensional locally adaptive non-hydrostatic model based on an extension of the SWE. Simply combining the ratio of surface elevation to the fluid depth with the norm of the horizontal velocity as the adaptivity criterion allows us to obtain similar results with the global model for all the test cases, saving up to nearly $40\%$ of the computational time, especially for cases that imitate tsunami waves. We plan to further investigate a more rigorous adaptivity criterion.

In the case of moving bottom-generated waves, our proposed model with both linear and quadratic pressure relations is not sufficient. This has already been observed in previous studies of linear profiles, suggesting that a multilayer model is more appropriate. Adapting the locally adaptive model to multilayer models should be relatively straightforward when it is solved with a projection method, which is the case in some previous studies. Hence, investigating a multilayer model with a quadratic relation is of interest for further development. However, the ability to achieve results close to the global model in less computational time remains the main advantage of this work.

\section*{Acknowledgements}
The authors acknowledge the support of the Deutsche Forschungsgemeinschaft (DFG, German Research Foundation) within the Research Training Group GRK 2583 ``Modeling, Simulation and Optimization of Fluid Dynamic Applications''. J.B. additionally acknowledges funding by the DFG under Germany‘s Excellence Strategy – EXC 2037 ``CLICCS - Climate, Climatic Change, and Society'' – Project Number 390683824; as well as through the Collaborative Research Center TRR 181 ``Energy Transfers in Atmosphere and Ocean'' funded by the DFG - Project Number 274762653. Moreover, K.F. would like to thank Prof. F.X. Giraldo for providing the MATLAB code for the 2D elliptic PDEs accompanying his book and for many helpful discussions regarding the implementation and numerical aspects of the method. Additional thanks go to Dr. M. Bänsch for guidance on Amatos and for sharing experience with LDG.

\section*{Conflicts of Interest}
The authors declare no conflicts of interest.

\section*{Data Availability Statement}
The data that support the findings of this study are available from the corresponding author upon reasonable request.

\bibliography{WileyNJD-AMA}
\end{document}